\documentclass[utf8]{frontiersSCNS} % for Science, Engineering and Humanities and Social Sciences articles
\usepackage{url,hyperref,lineno,microtype,subcaption}
\usepackage[onehalfspacing]{setspace}

\usepackage{amsmath}
\usepackage{amssymb}

\hypersetup{
    colorlinks=true,
    citecolor=blue,
    linkcolor=blue,
    filecolor=magenta,      
    urlcolor=blue,
}

\graphicspath{ {Figures/} }

\def\keyFont{\fontsize{8}{11}\helveticabold }
\def\firstAuthorLast{Correia {et~al.}} %use et al only if is more than 1 author
\def\Authors{Rion Brattig Correia\,$^{1,2,3}$, Alexander J.\ Gates\,$^{4}$, Xuan Wang\,$^{1}$ and Luis M.\ Rocha\,$^{1,3,*}$}
\def\Address{$^{1}$School of Informatics, Computing, and Engineering Indiana University, Bloomington, IN, USA \\
$^{2}$CAPES Foundation, Ministry of Education of Brazil, Brasília, DF, Brazil\\
$^{3}$Instituto Gulbenkian de Ciência, Oeiras, Portugal \\
$^{4}$Center for Complex Networks Research, Northeastern University, Boston, MA, USA}
\def\corrAuthor{901 E. 10\textsuperscript{th} St. Bloomington, IN 47408}
\def\corrEmail{rocha@indiana.edu}

\begin{document}
\onecolumn
\firstpage{1}

\title[\textbf{CANA}]{\textbf{CANA}: A python package for quantifying control and canalization in Boolean Networks} 

\author[\firstAuthorLast ]{\Authors} %This field will be automatically populated
\address{} %This field will be automatically populated
\correspondance{} %This field will be automatically populated

\extraAuth{}% If there are more than 1 corresponding author, comment this line and uncomment the next one.
%\extraAuth{corresponding Author2 \\ Laboratory X2, Institute X2, Department X2, Organization X2, Street X2, City X2 , State XX2 (only USA, Canada and Australia), Zip Code2, X2 Country X2, email2@uni2.edu}

\maketitle

\begin{abstract}

Logical models offer a simple but powerful means to understand the complex dynamics of biochemical regulation, without the need to estimate kinetic parameters. However, even simple automata components can lead to collective dynamics that are computationally intractable when aggregated into networks. 
In previous work we demonstrated that automata network models of biochemical regulation are highly canalizing, whereby many variable states and their groupings are redundant \citep{MarquesPita:2013}. 
The precise charting and measurement of such canalization simplifies these models, making even very large networks amenable to analysis. 
Moreover, canalization plays an important role in the control, robustness, modularity and criticality of Boolean network dynamics, especially those used to model biochemical regulation \citep{Gates:2016,GatesManickaMarquesPitaRocha:2016, Manicka:2017}. 
Here we describe a new publicly-available Python package that provides the necessary tools to extract, measure, and visualize canalizing redundancy present in Boolean network models.
It extracts the pathways most effective in controlling dynamics in these models, including their \textit{effective graph} and \textit{dynamics canalizing map}, as well as other tools to uncover minimum sets of control variables.

\tiny
 \keyFont{ \section{Keywords:} Boolean Networks, Automata, Canalization, Python package, biochemical regulation, Logical modeling, Network dynamics, Complex systems }
 %All article types: you may provide up to 8 keywords; at least 5 are mandatory.
\end{abstract}

\section{A Tool to study redundancy and control in Boolean networks}
\label{ch:intro}

% intro to logical models
Mathematical and computational modelling of biological networks promises to uncover the fundamental principles of living systems in an integrative manner \citep{Ideker:2017,Iyengar:2009}.
In particular, Boolean Networks (BN), a class of logical dynamical systems, provide an effective framework to capture the dynamics of interconnected biological systems without the need for detailed kinetic parameters \citep{Assmann:2009,Bornholdt:2005}.
BN have been used to model and predict
    biochemical regulation in genetic networks \citep{Li:2004},
    cell signalling \citep{Helikar:2008},
    chemical reactions in metabolic networks \citep{Chechik:2008},
    anticancer drug response \citep{Choi:2017},
    action potentials in neural networks \citep{Kurten:1988},
    and many other dynamical systems involved in biomedical complexity \citep{Albert:2003}.

Two reasons contribute to the success of BN models: (i) the reduction of complex multivariate dynamics to a graph revealing  the organization and constraints of the topology of interactions in biological systems, and (ii) a coarse-grained treatment of dynamics that facilitates predictions of limiting behavior and robustness \citep{Bornholdt:2008}.
However, more than understanding the organization of complex biological systems, we need to derive control strategies that allow us, for example, to intervene on a diseased cell \citep{Zhang:2008}, or revert a mature cell to a pluripotent state \citep{Wang:2011}.
Recently, several mathematical tools were developed to enhance our understanding of BN control by removing redundant pathways, identifying key dynamic modules \citep{MarquesPita:2013}, and characterizing critical driver variables \citep{Gates:2016}.

Here we present CANA\footnote{\textbf{CANA}lization: Redundancy \& Control in Boolean Networks. For documentation and tutorials see \url{github.com/rionbr/CANA}.}, a python package to study redundancy and control in BN models of biochemical dynamics \citep{CANA:github}.
It provides a simple interface to access computational tools for three important aspects of BN analysis and prediction:
\begin{enumerate}
    \item \textbf{Dynamics.} Python classes are included to enumerate all \emph{attractors} and calculate the full \emph{state transition graph} (STG) of BN, as described in \S\ref{ch:notation}.
    \item \textbf{Canalization.} The redundancy properties of automata functions have been characterized as a form of canalization \citep{kauffman:1984}, particularly when used to model  dynamical interactions in models of genetic regulation and biochemical signalling \citep{Reichhardt:2007,Kauffman:2004,MarquesPita:2013}. At the level of individual Boolean transition functions (network nodes), canalization is observed when not all inputs are necessary to determine a state transition (see  \S\ref{ch:redundancy} for formal definition). CANA can be used to calculate all measures of canalization that derive from removing dynamical redundancy via two-symbol schemata re-description \citep{MarquesPita:2013}: \emph{effective connectivity}, \emph{input redundancy} and \emph{input symmetry}. At the network level, CANA also calculates the \emph{effective graph}, %\citep{Gates:2018}, 
    a weighted and directed graph whose edge weights denote their effective contribution to node transitions, as well as the \emph{dynamics canalizing map}, 
    %\citep{MarquesPita:2013}, 
    a parsimonious representation of the necessary and sufficient state transitions that define the entire dynamics of BN. All canalization measures and network representations are applicable to synchronous and asynchronous BN models, as described in \S\ref{ch:redundancy}.
    \item \textbf{Control.} From a subset of driver variables---nodes that act as the loci of control interventions---CANA computes the \emph{controlled state transition graph} (CSTG), as well as the \emph{controlled attractor graph} (CAG) capturing all controlled transitions between attractors possible via driver variable interventions \citep{Gates:2016}. CANA also computes measures of controllability that depend on the CSTG and CAG: \emph{mean fraction of reachable configurations}, \emph{mean fraction of controlled configurations}, and \emph{mean fraction of reachable attractors}, as described in \S\ref{ch:control}. Currently, control analysis in CANA is applicable only to synchronous BN models.
\end{enumerate}
Here we demonstrate the full functionality of the CANA package using the BN model of floral organ development in the flowering plant \emph{Arabidopsis Thaliana} \citep{Chaos:2006}.
Additionally, we provide an interface between CANA and the \textit{Cell Collective} \citep{Helikar:2012}, allowing for an extensive analysis of control and canalization in complex biological systems.

The CANA package fills a key void in the available library of computational software to analyze Boolean Network models. 
Existing software falls into two categories: either they are designed to reverse engineer BN models from biological experimental data, or they focus on simulating BN dynamics.
Examples of the first category include the \emph{CellNetOptimizer} which creates BN from high-throughput biochemical data \citep{Terfve:2012}, and the \emph{Dynamic Deterministic Effects Propagation Networks} (DDEPN) package which reconstructs signalling networks based from time-course experimental data \citep{Bender:2010}.
The second category of BN simulation packages is best exemplified by \emph{BooleanNet}, a python package that simulates both synchronous and asynchronous dynamics \citep{IAlbert:2008}, and PANET, a Cytoscape plugin that quantifies the robustness of BN models \citep{Trinh:2014}. \emph{The Cell Collective}, a collaborative platform and intuitive visual interface to share and build BN models, can also be used to simulate BN dynamics \citep{Helikar:2012}.
The CANA package expands the set of available tools of the second category, by providing Python classes to calculate measures and visualizations of canalization (dynamical redundancy) and control of BN models, as detailed below.
CANA is designed as a toolbox for both computational and experimental system biologists. It enables the simplification of BN models and testing of network control algorithms, thus prioritizing biochemical variables more likely to be relevant for specific biological questions (e.g. genes controlling cell fate), and ideal candidates for knockout experiments.
%

% Formal definition (notation from MarquesPita:2013 & Gates:2016}
\section{Boolean Network Representation and Dynamics}
\label{ch:notation}

A \emph{Boolean automaton} is a binary variable, $x \in \{0,1\}$, whose state is updated in discrete time-steps, $t$, according to a deterministic \emph{Boolean state-transition function} of $k$ inputs: $x^{t+1} = f(x_{1}^t,...,x_{k}^t)$. 
The state-transition function, $f:\{0,1\}^k \to \{0,1\}$, is defined by a \emph{look-up (truth) table} (LUT), $F \equiv \{ f_{\alpha} : \alpha=1,...,2^k \}$, with one entry for each of the $2^k$ combinations of input states and a mapping to the automaton's next state (transition or output), $x^{t+1}$ (Figure \ref{fig:1}A).
In CANA, a Boolean automaton---a python class denoted \emph{BooleanNode}---is instantiated from the list of transitions that define its LUT.

A \emph{Boolean Network} is a graph $\mathcal{B} \equiv (X,C)$, where $X$ is a set of $N$ Boolean automata \emph{nodes} $x_i \in X, i=1,...,N$ and $C$ is a set of directed edges $c_{ji} \in C : x_i, x_j \in X$ that represent the interaction network, denoting that automaton $x_j$ is an input to automaton $x_i$, as computed by $F_i$. 
The set of inputs for automaton $x_i$ is denoted by $X_i = \{ x_j \in X : c_{ji} \in C \}$, and its cardinality, $k_i = |X_i|$, is the \emph{in-degree} of node $x_i$.
At any given time $t$, $\mathcal{B}$ is in a specific configuration of automata states, $\boldsymbol{x}^t = \langle x_1^t,x_2^t,...,x_N^t \rangle$, where we use the terms \emph{state} for individual automata ($x_i^{t}$) and \emph{configuration} ($\boldsymbol{x}^t$) for the collection of states of all automata of the BN at time $t$, i.e.\ the collective network state.
The set of all possible network configurations is denoted by $\mathcal{X} \equiv \{0,1\}^N$, where $|\mathcal{X}| = 2^N$.
The dynamics of $\mathcal{B}$ unfolds from an initial configuration, $\boldsymbol{x}^0$, by a \emph{synchronous} update policy in which all automata transition to the next state at each time step. 
The complete dynamical behavior of the system for all initial conditions is captured by the \emph{state-transition graph} (STG), $\mathcal{G} \equiv \text{STG}(\mathcal{B}) = (\mathcal{X}, \mathcal{T})$, where each node is a configuration $\boldsymbol{x}_{\alpha} \in \mathcal{X}$, and an edge $T_{\alpha,\beta} \in \mathcal{T}$ denotes that a BN in configuration $\boldsymbol{x}_{\alpha}$ at time $t$ will be in configuration $\boldsymbol{x}_{\beta}$ at time $t+1$.
Under deterministic dynamics, only a single transition edge $T_{\alpha,\beta}$ is allowed out of every configuration node $\boldsymbol{x}_{\alpha}$.
Configurations that repeat, such that $\boldsymbol{x}_{\alpha}^{t+\mu} = \boldsymbol{x}_{\beta}^{t}$, are known as \emph{attractors} and differentiated as \emph{fixed-point} attractors when $\mu=1$, and \emph{limit cycles} when $\mu>1$, respectively.
Because $\mathcal{G}$ is finite, it contains at least one attractor, as some configuration or limit cycle must repeat in time \citep{Wuensche:1998}.

In CANA, a python class named \emph{BooleanNetwork} represents a BN, and is instantiated from a dictionary containing the transition functions (LUT) of all its constituent  automata nodes, or loaded from a file. 
We also provide several predefined example BN models that can be directly loaded: the \emph{Arabidopsis Thaliana} gene regulatory network (GRN) of flowering patterns \citep{Chaos:2006}, a simplified version of the segment polarity GRN of \emph{Drosophila Melanogaster} \citep{Albert:2003}, the \emph{Budding Yeast} cell-cycle regulatory network \citep{Li:2004}, and the BN motifs analyzed in \cite{Gates:2016}.
Beyond the aforementioned networks, our current release also incorporates all publicly available networks in the Cell Collective repository \citep{Helikar:2012}. These were loaded from the Cell Collective API and converted into truth tables that can be read by CANA
\footnote{Future releases will provide a direct link to the Cell Collective API for conversion of Cell Collective models. Currently, models are converted to .CNET (truth table) format,and subsequently imported to CANA.}.
CANA has two built-in methods available to compute network dynamics: for relatively small BN ($N<30$) the full state-space can be computed, whereas for larger BN, CANA uses a Boolean satisfiability (SAT) based algorithm, capable of enumerating all attractors in a BN with thousands of variables \citep{Dubrova:2011}.

\section{Canalization}
\label{ch:redundancy}

\begin{figure}[h!]
    \begin{center}
        \includegraphics[width=\textwidth]{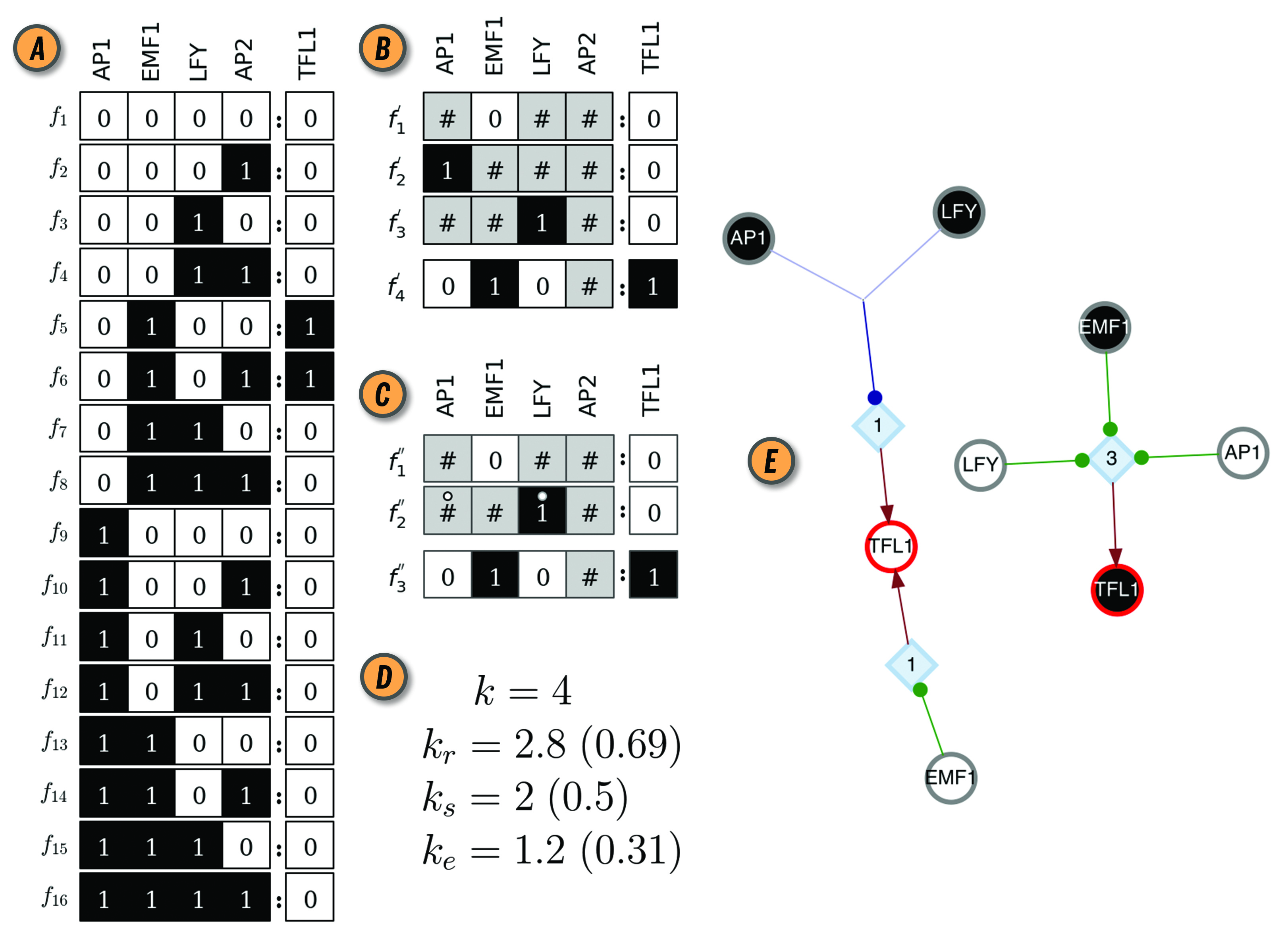}
    \end{center}
    \caption{
        CANA analysis of Boolean automaton for TFL1 in the BN model of the floral organ arrangement in the flowering plant \emph{Arabidopsis Thaliana}.
        \textbf{(A)} Look-up-table (LUT).
        \textbf{(B)} Wildcard schema redescription, $F'$(TFL1).
            Wildcards are denoted by gray states.
            As an example, schema $f'_4$ redescribes the subset of LUT entries $\Upsilon_4 \equiv \{f_5,f_6\}$, where $i_4 = \text{AP2}$ can be either \textit{on} or \textit{off}.
        \textbf{(C)} Two-symbol schema redescription, $F''$(TFL1).
            Permutation of the inputs marked with the position-free symbol ($\circ$) in any schema of $F''$(TFL1) result in a wildcard schema in $F'$(TFL1). 
            For example, $f_{2}''$ redescribes $\Theta_2' \equiv \{ f_2',f_3' \}$.
        \textbf{(D)} In-degree ($k$), input redundancy ($k_r$), input symmetry ($k_s$), and effective connectivity ($k_e$) of TFL1 automaton. Values in parenthesis are the respective (relative) measures normalized by $k$, used for comparisons between automata with different number of inputs.
        \textbf{(E)} Canalizing Map (CM) of automaton TFL1, with its two possible states, TFL1 $\in \{0,1\}$, shown as circles with red contour; white (black) fill color denotes state 0 (1).
        Input variables and their respective state are also shown as circles (\textit{s-units}) with the same color criterion, and link to \textit{t-units} shown as blue diamonds with corresponding threshold value inside; thus, TFL1 requires 3 input conditions (LFY=0 $\land$ EMF1=1 $\land$ AP1=0) to turn on (TLF1=1), but only one (EMF1=0 $\lor$ AP1 =1 $\lor$ LFY = 1) to turn off (TLF1=0); $\land$ and $\lor$ denote the logical conjunction (and) and disjunction (or), respectively. 
        Network rendering generated with Graphviz \citep{Graphviz:2002}.
    }
    \label{fig:1}
\end{figure}

Important insights about BN dynamics are gained by observing that not all inputs to an automaton are equally important for determining its state transitions, a concept known as \emph{canalization} \citep{Reichhardt:2007}.
Originally, the term was proposed by \cite{waddington:1942} and subsequently refined to characterize the buffering of genetic and epigenetic perturbations leading to the stability of phenotypic traits \citep{siegal:2002,masel:2007,tusscher:2009}.
Understanding how canalization occurs in a given BN model allows us to uncover and remove redundancy present in the pathways that control its dynamics.
In CANA, we follow \cite{MarquesPita:2013} by quantifying canalization through the logical \textit{redundancy} present in automata.
Specifically, we use the Quine-McCluskey Boolean minimization algorithm \citep{Quine:1955} to identify those inputs of an automaton which are redundant given the state of its other inputs, thus reducing its LUT to a set of \emph{prime implicants}. 
The prime implicants are in turn combined to create wildcard schemata, $F' \equiv \{ f'_{\upsilon} \}$, in which the \emph{wildcard} or ``Don't care''symbol, \#, (also represented graphically in grey) denotes an input whose state is redundant given the state of other necessary input states.
%$\Upsilon_{\nu} \equiv \{ f_{\alpha} : f_{\alpha} \mapsto f'_{\nu} \}$ 
%
In this process, the original LUT $F$ (Figure \ref{fig:1}A) is redescribed by a more compressed set of schemata $F'$ (Figure \ref{fig:1}B).
Every wildcard schema  $f'_{\upsilon} \in F'$ redescribes a subset of entries in the original LUT,
denoted by $\Upsilon_{\upsilon} \equiv \{f_{\alpha}: f_{\alpha}
\rightarrowtail f'_{\upsilon}\} \subseteq F$; $\rightarrowtail$ means `is
redescribed by'.
Finally, CANA also calculates the \emph{two-symbol schemata} redescription, $F'' \equiv \{ f''_{\theta} \}$, whereby in addition to the wildcard symbol, a \emph{position-free} symbol, $\circ$, further captures \emph{permutation redundancy} (i.e.\ group-symmetry): subsets of inputs whose states can permute without affecting the automaton's state (Figure \ref{fig:1}C).
Every two-symbol schema $f''_{\theta} \in F''$ redescribes a set $\Theta_{\theta} \equiv
\{f_{\alpha}: f_\alpha \rightarrowtail f''_{\theta}\} \subseteq F$ of LUT entries
of automaton $x$.

Several measures of canalization present in the LUT of an automaton are also defined in CANA, and can be accessed by function calls to both the \emph{BooleanNode} and \emph{BooleanNetwork} classes.
\emph{Input redundancy}, $k_r(x)$, measures the number of inputs that on average are not needed to compute the state of automaton $x$. This is measured by tallying the mean number of wildcard symbols present in the set of schemata $F'(x)$ or $F''(x)$ that redescribe the LUT $F(x)$ (eq. \ref{eq:canalization}).
\emph{Effective connectivity}, $k_e$, is a complementary measure of $k_r (x)$ yielding the number of inputs that are on average necessary to compute the automaton's state (eq. \ref{eq:canalization}). Whereas $k(x)$ is the number of inputs to automaton $x$ present in the BN, $k_e(x)$ is the minimum number of such inputs that are on average necessary to determine the state of $x$---its effective connectivity or degree. 
Similarly, \emph{input symmetry}, $k_s (x)$, is the mean number of inputs that can permute without effect on the state of $x$. It is measured by tallying the mean number of position-free symbols present in $F''(x)$ (eq.\ \ref{eq:canalization}):
\begin{equation}
\label{eq:canalization}
    k_r(x) = \frac{ \sum\limits_{f_{\alpha} \in F} \; \underset{\upsilon:f_{\alpha} \in \Upsilon_{\upsilon}}{\max} \; \big( n^{\#}_{\upsilon} \big) }{ |F| }
    \quad , \quad 
    k_e(x) = k(x) - k_r(x)
    \quad , \quad
    k_s(x) = \frac{ \sum\limits_{f_{\alpha} \in F} \; \underset{\theta:f_{\alpha} \in \Theta_{\theta}}{\max} \; \big( n^{\circ}_{\theta} \big) }{ |F| }
    \quad 
\end{equation}
\noindent where $n^{\#}_{\upsilon}$ and $n^{\circ}_{\theta}$ are the number of inputs with a \#  or $\circ$ in schema $f'_{\upsilon}$ or $f''_{\theta}$, respectively\footnote{$k_r$ and $k_e$ can be computed on either set of schemata $F'$ (as in eq. \ref{eq:canalization}) or $F''$ (as in \citealt{MarquesPita:2013}), yielding the same result; $k_s$ must be computed on $F''$.}.
Figure \ref{fig:1}D shows the values of these measures for the LUT of the TFL1 gene in the \textit{thaliana} GRN model.
Additional algorithmic details of the two forms of canalization, as well as their importance to study control, robustness, and modularity of BN models of biochemical regulation, are presented in \cite{MarquesPita:2013}. Next we introduce new per-input measures of canalization as well as the effective graph, which CANA also computes.

Most automata contain redundancy of one or both of the two forms of canalization; only the two parity functions for any $k$ have $k_r=0$ (e.g. the $XOR$ function and its negation for $k=2$), and even those can have $k_s>0$.
Therefore, the original interaction graph of a BN tends to have much redundancy and does not capture how automata truly influence one another in a BN. 
To formalize this idea, the CANA package computes an \textit{effective graph}, $\mathcal{E} \equiv (X,E)$, where $X$ is as in \S \ref{ch:notation} and $E$ is a set of weighted directed edges $e_{ji} \in [0,1] \forall x_i, x_j \in X$ denoting  the \textit{effectiveness} of automaton $x_j$ in determining the truth value of automaton $x_i$, and computed via eq. \ref{eq:per_input_canalization}.
Specifically, we define per-input measures of canalization for \textit{redundancy}, \textit{effectiveness}, and \textit{symmetry}, respectively:

\begin{equation}
\label{eq:per_input_canalization}
    r_{ji} = \frac{ \sum\limits_{f_{\alpha} \in F_i} \; \underset{\upsilon:f_{\alpha} \in \Upsilon_{\upsilon}^i}{\textrm{avg}} \; \big( j \rightarrowtail \# \big)_{\upsilon} }{ |F_i| }
    %\
    \quad , \quad 
    e_{ji} = 1 - r_{ji}
    \quad , \quad
    s_{ji} = \frac{ \sum\limits_{f_{\alpha} \in F_i} \; \underset{\theta:f_{\alpha} \in \Theta_{\theta}^i}{\textrm{avg}} \; \big(j  \rightarrowtail \circ \big)_{\theta} }{ |F_i|}
    \quad 
\end{equation}

\noindent where $( j \rightarrowtail \# )_{\upsilon}$ is a logical condition that assumes the truth value $1(0)$ if input $j$ is (not) a wildcard in schema $f'_{\upsilon}$, and similarly for $(j  \rightarrowtail \circ)_{\theta}$ for a position-free symbol in schema $f''_{\theta}$; $\textrm{avg}$ is the average operator.
Naturally, $k_r(x_i) = \sum_{j} r_{ji}$, $k_e(x_i) = \sum_{j} e_{ji}$, and $k_s(x_i) = \sum_{j} s_{ji}$.

\begin{figure}[h!]
    \begin{center}
        \includegraphics[width=.80\textwidth]{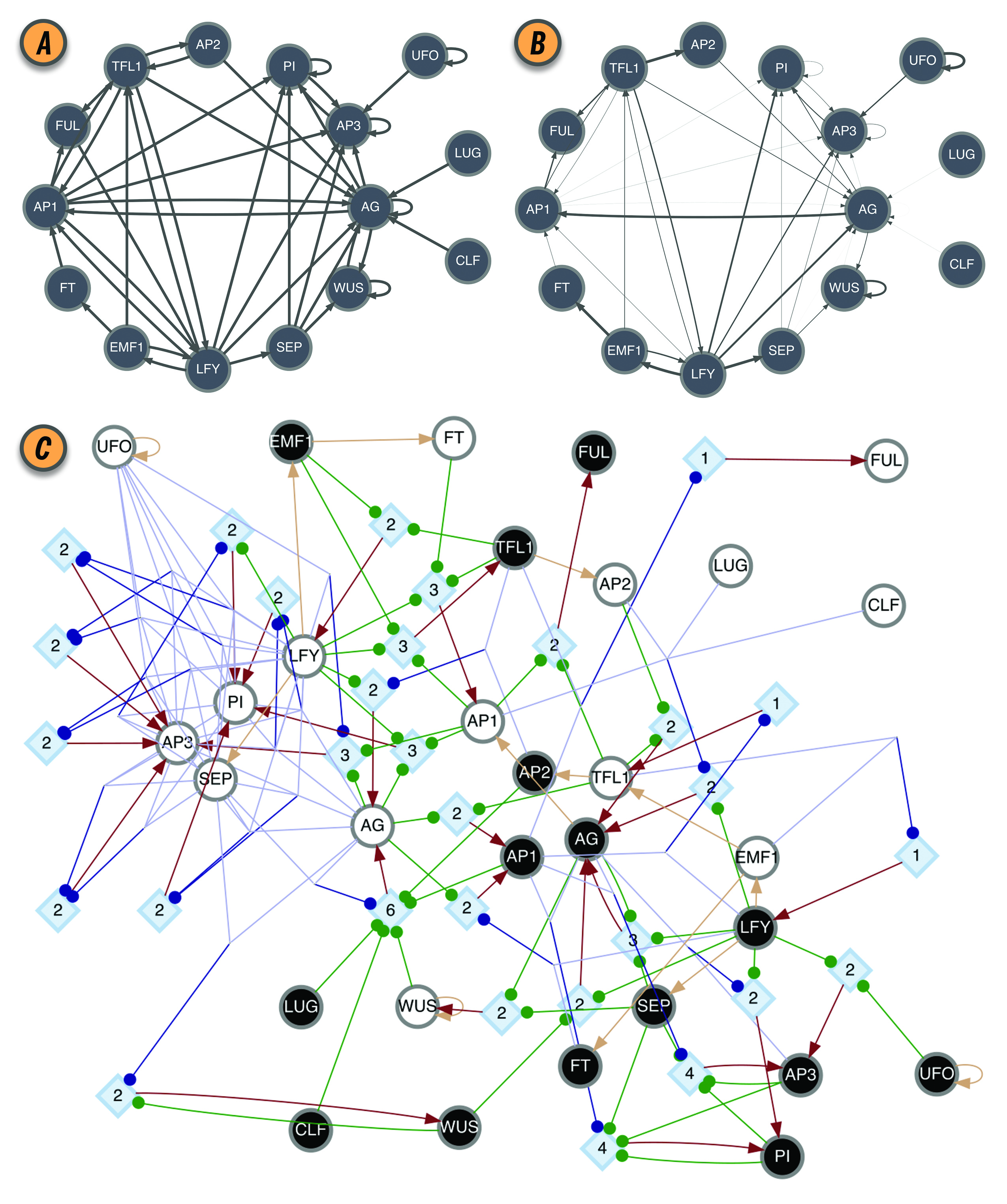}
    \end{center}
    \caption{
        BN model of the floral organ arrangement in the flowering plant \emph{Arabidopsis Thaliana}.
        \textbf{(A)} Interaction graph $C$.
        \textbf{(B)} Effective graph $E$, where edge weights denote $e_{ji}$ (eq. \ref{eq:per_input_canalization}).
            Some edges, originally in $C$, are completely removed in $E$ (e.g., AG$\to$AG, AP1$\to$AG, and AP2$\to$TFL1). Others, have very small effectiveness (e.g., AP1$\to$PI and CLF$\to$AG).
        \textbf{(C)} Dynamics Canalization Map (DCM) representing the entire logic of interactions after removal of redundancy.
            Original BN automata nodes appear twice in the DCM, once for each Boolean truth value and denoted as \textit{s-unit}, white (0) or black (1) circles.
            When \textit{s-units} are determined by another single \textit{s-unit} they are connected with a beige directed edge---a simplification to avoid the rendering of a \textit{t-unit} with a threshold of one. All other variable state determinations occur via \textit{t-units} with larger threshold values. Red edges represent outputs from \textit{t-units} to \textit{s-units}: a state determination of the receiving \textit{s-unit}, after the logical condition of the \textit{t-unit} is met.All other (blue or green) edges denote inputs from \textit{s-units} to \textit{t-units}, that is, the sufficient conditions for a state determination. Blue edges denote group disjunction constraints, whereby conditions captured by \textit{s-units} can merge because any one of the merging conditions is sufficient (e.g., $\left(\text{TFL}=0\lor\text{EMF1}=0\right)\to\text{LFY}=1$). Green edges denote independent and necessary conditions. Directed edges into \textit{s-units} are denoted by arrows, while directed edges into \textit{t-units} are denoted by small circles.
        Network rendering by Graphviz \citep{Graphviz:2002}.
    }
    \label{fig:2}
\end{figure}

The effective graph was shown to be important in predicting the controllability of BN \citep{Gates:2016}. Furthermore, the mean $k_e$ of BN (the mean in-degree of the effective graph) is a better predictor of criticality than the in-degree of the original interaction graph \citep{Manicka:2017}, improving the existing theory for predicting criticality in BN \citep{Aldana:2003}. 
Those results suggest that Natural Selection can select for canalization, thereby enhancing the stability and controllability of networks with high connectivity, that would otherwise exist in the chaotic regime \citep{GatesManickaMarquesPitaRocha:2016,Manicka:2017}.
%\citep{ManickaMarquesRocha:2018}.
%
As an example, the interaction and effective graphs of the \textit{Thaliana} GRN BN model, as computed by CANA, are shown in Figure \ref{fig:2}A-B, demonstrating that much redundancy exists in the original model. 
The most extreme case of redundancy occurs when an input from $x_j$ to automaton $x_i$ exists in the original interaction graph $C$, $c_{ji}=1$, but not in the effective graph $\mathcal{E}$, $e_{ji}=0$, because it is fully redundant and does not affect the automaton's transition (see several such cases in Figure \ref{fig:2}A-B).

The canalizing logic of an automaton provided by the schemata set $F''$, can also be represented as a \cite{McCulloch:1943} threshold network, named  a \emph{Canalizing Map} (CM) in \cite{MarquesPita:2013}. Figure \ref{fig:1}E depicts the CM for the TFL1 gene.
It consists of two types of nodes: \textit{state units} (\textit{s-unit}, denoted by circles), which represent automata in one of the Boolean truth values ($x_i = 0$, white, or $x_i = 1$, black), and \textit{threshold units} (\textit{t-unit}, denoted by diamonds), which implement a numerical threshold condition on its inputs. 
When the CM of all automata of a BN are linked, we obtain the \textit{Dynamics Canalization Map} (DCM), as shown in Figure \ref{fig:2}C for the \textit{Thaliana} GRN.
Directed fibers connect nodes and propagate an activation pulse; fibres can merge and split, but each end-point always contributes one pulse to an s-unit.
The DCM is a highly parsimonious representation of the dynamics of a BN. It contains only necessary information about how (canalizing) control signals determine network dynamics. It enables inferences about control, modularity and robustness to be made about the collective (macro-level) dynamics of BN \citep{MarquesPita:2013}. Because it is assembled using solely the micro-level canalizing logic of individual automata, its computation scales linearly with the number of nodes of the network, and thus it can be computed for very large networks.
The computational bottleneck can only be the number of inputs ($k$) to a particular automaton, since the Quine–McCluskey algorithm grows exponentially with the number of variables. Functions with a large number of variables have to be minimized with heuristic methods such as Espresso \citep{Brayton:1984}.
Because all measures of canalization, as well as the effective graph and the DCM, derive from removing dynamical redundancy at the level of individual automata, they are independent from the updating regime chosen for the network. In other words, the canalization analysis are applicable to synchronous and asynchronous BN models.

\section{Control}
\label{ch:control}

The discovery of control strategies in BN models is a central problem in systems biology; theoretical insights about controllability can enhance experimental turnover by focusing experimental interventions on genes and proteins more likely to result in the desired phenotype output.
It is well known that when the set of automata nodes $X$ of a BN is large, enumeration of all configurations $\boldsymbol{x} \in \mathcal{X}$ of its STG becomes difficult, making the controllability of deterministic BN an NP-hard problem \citep{Akutsu:2007}.
Thus control methodologies which leverage the interaction graph or remove the redundancy in canalizing automata are highly desirable, since they can greatly simplify BN complexity.

CANA contains Python functions designed to provide a testbed for the development of BN control strategies, and to investigate the interplay between canalization, control, and other dynamics properties.
Specifically, we study the control exerted on the dynamics of a BN, $\mathcal{B} = (X, C)$, by a subset of \emph{driver variables} $D \subseteq X$---a subset of automata nodes of $\mathcal{B}$.
Control \emph{interventions} are realized by instantaneous bit-flip perturbations to the state of the variables in $D$ \citep{Willadsen:2007}.
This results in a  \emph{controlled state transition graph}, $\text{CSTG}(\mathcal{B})\equiv \mathcal{G}_D \equiv (\mathcal{X}, \mathcal{T} \cup \mathcal{T}_D)$, which is an extension of the STG that captures all possible trajectories due to controlled interventions on $D$ \citep{Gates:2016}.
The additional edges $\mathcal{T}_D$ denote transitions from every configuration to a set of $2^{|D|}-1$ configurations in the STG, which are reachable given the bit-flip perturbations of the driver variables.
A BN is controllable when every configuration is reachable from every other configuration in $\mathcal{G}_D$ \citep{Sontag:1998}, a condition equivalent to requiring that the CSTG $\mathcal{G}_D$ be strongly connected.

CANA computes the CSTG of $\mathcal{B}$ given a driver set $D$, which in turn is used to calculate the \emph{mean fraction of reachable configurations}, $\overline{R}_D $, and the \emph{mean fraction of controlled configurations}, $\overline{C}_D$, \citep{Gates:2016}:
\begin{equation}
	\label{eq:frac_reached}
	\overline{R}_D = \frac{1}{2^{N}} \sum_{\boldsymbol{x}_{\alpha} \in \mathcal{X}} r(\mathcal{G}_D,\boldsymbol{x}_{\alpha})
	\quad , \quad 
	\overline{C}_D =\overline{R}_D - \overline{R}_{\emptyset}
	\quad .
\end{equation}
\noindent where, for each configuration $\boldsymbol{x}_{\alpha}$,  $r(\mathcal{G}_D, \boldsymbol{x}_{\alpha})$ is the \emph{fraction of reachable configurations}, defined as the number of other configurations $\mathbf{X}_\beta$ lying on all directed paths from $\boldsymbol{x}_{\alpha}$, normalized by the total number of other configurations $2^{N-1}$.
Similarly, the \emph{fraction of controlled configurations} counts the number of new configurations that are reachable due to interventions to $D$, but were not originally reachable in the STG: $c(\mathcal{G}_D, \boldsymbol{x}_{\alpha}) = r(\mathcal{G}_D, \boldsymbol{x}_{\alpha}) - r(\mathcal{G}, \boldsymbol{x}_{\alpha})$.
When a BN is fully controlled by $D$, $\overline{R}_D = 1.0$, but for partially controlled BNs $\overline{R}_D \in [0.0,1.0)$; note that $\overline{C}_D \leq \overline{R}_D$.

\begin{figure}[h!]
    \begin{center}
        \includegraphics[width=\textwidth]{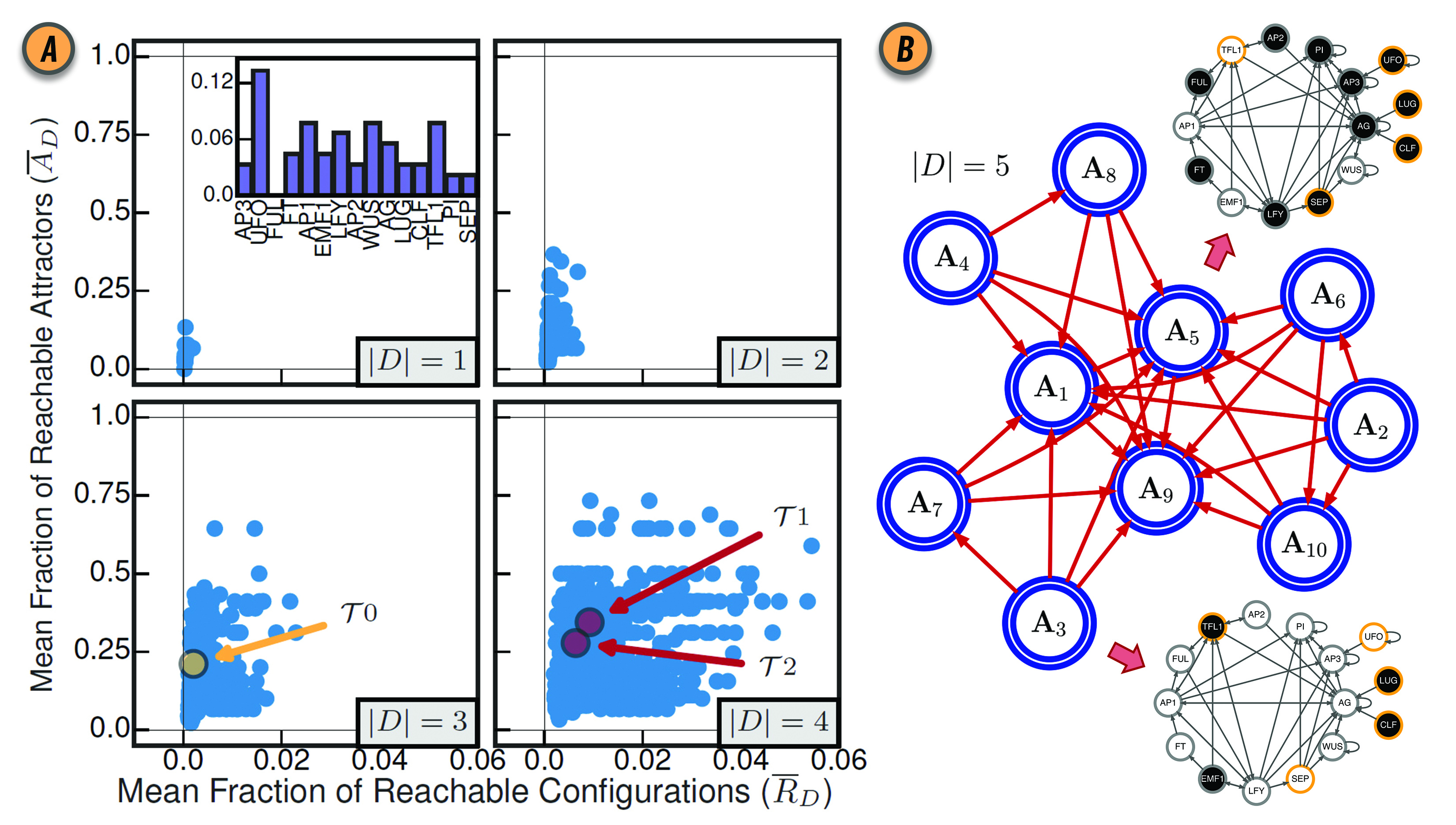}
    \end{center}
    \caption{
        \textbf{(A)}
            Control of the BN model of gene regulation involved in the floral organ development in the \emph{Arabidopsis thaliana} plant for all driver variable subsets of size $|D|=1$, $|D|=2$, $|D|=3$ and $|D|=4$. (\textbf{inset}) The mean fraction of reachable attractors $\overline{A}_D$ for each singleton drive variable set. The driver variable subsets predicted by structural controllability to fully control the network are highlighted in red and labeled $\mathcal{T}1$ and $\mathcal{T}2$. The three variable subset with all three root variables is highlighted in yellow and labeled $\mathcal{T}0$.
            Reproduced from \cite{Gates:2016} under Creative Commons Attribution 4.0 International (CC BY 4.0) license.
        \textbf{(B)}
            The CAG with driver variables $D=\{\text{UFO},\text{LUG},\text{CLF},\text{SEP},\text{TFL1}\}$. Each large blue node $\mathbf{A}_{1},\ldots,\mathbf{A}_{10}$ represents an attractor of the network dynamics.
            The BN configurations for steady-state attractors $\mathbf{A}_3$ and $\mathbf{A}_5$ are shown as interaction graphs with node variables colored white or black for states $x_i=0$ and $x_i=1$, respectively; driver variables are shown with a yellow contour.
    }
    \label{fig:3}
\end{figure}

In Systems Biology applications, typically only the attractors of BN are meaningful configurations, used to represent different cell types \citep{Kauffman:1993, Muller:2011, Kauffman:1969}, diseased or normal conditions \citep{Zhang:2008}, and wild-type or mutant phenotypes \citep{Albert:2003}.
In this context, a more relevant control measure is the extent to which driver variables can steer dynamics from attractor to attractor.
To quantify such control, CANA computes the \emph{controlled attractor graph} (CAG) of a BN $\mathcal{B}: \mathcal{C}_D = (\mathcal{A},\mathcal{Z}_D)$.
The nodes of this graph, $\mathbf{A}_{\kappa} \in \mathcal{A}$, represent an attractor of $\mathcal{B}$, and each edge $z_{\kappa\gamma} \in \mathcal{Z}_D$, denotes the existence of at least one path from attractor $\mathbf{A}_{\kappa}$ to attractor $\mathbf{A}_{\gamma}$ in the CSTG $\mathcal{G}_D$ (Figure \ref{fig:3}B).
The \emph{mean fraction of reachable attractors} is then given by
\begin{equation}
\label{eq:frac_attract_reach}
	\overline{A}_D =\frac{1}{|\mathcal{A}|} \sum_{\mathbf{A}_\kappa \in \mathcal{A}} r(\mathcal{C}_D,\mathbf{A}_{\kappa})
\end{equation}
where $\kappa=1 \ldots |\mathcal{A}|$ \citep{Gates:2016}.
Since this notion of control depends only on the enumeration of attractors, CANA can leverage a SAT-based bounded model algorithm to quantify the mean fraction of reachable attractors in a BN with thousands of variables \citep{Dubrova:2011}. 
Figure \ref{fig:3}A shows the values of $\overline{R}_D$ and $\overline{A}_D$ for various sizes of driver sets $D$ in the \textit{Thaliana} GRN.

Finally, CANA also provides the functionality to approximate the minimal driver variable subset using two prominent network control methodologies: \emph{Structural Controlability} (SC) \citep{Lin:1974, Liu:2011} and \emph{Minimum Dominating Set} (MDS) \citep{Nacher:2012, Nacher:2013}.

\section{Summary and Conclusion}

We presented a novel, open-source and publicly-available software platform that integrates the analytic methodology used to study canalization in automata network dynamics.
This methodology can now be used by others to simplify large automata networks, especially those in models of biochemical regulation dynamics. 
In addition to the extraction and visualization of specific effective pathways that regulate key phenotypic outcomes in a sea of redundant interaction, CANA includes functionality to measure canalization, uncover control variables, and study dynamical modularity, robustness, and criticality.
We hope that the consolidation of redundancy and control algorithms into one package encourages other researchers to build upon our work on canalization, thus adding additional algorithms to CANA.

%%%%%%%%%%%%%%%%%%%%%%%%%%%%%%%%%%%%%%%%%%%%%%%%%%%
% End Matter
%%%%%%%%%%%%%%%%%%%%%%%%%%%%%%%%%%%%%%%%%%%%%%%%%%%

\section*{Conflict of Interest Statement}
%All financial, commercial or other relationships that might be perceived by the academic community as representing a potential conflict of interest must be disclosed. If no such relationship exists, authors will be asked to confirm the following statement: 

The authors declare that the research was conducted in the absence of any commercial or financial relationships that could be construed as a potential conflict of interest.

\section*{Author Contributions}

RCB, AJG, XW contributed to the CANA package. LMR developed the per-input measures of canalization and the effective graph formulation.  RCB, AJG, and LMR wrote the manuscript. 

\section*{Funding}
%Who gave us the moneys?
RBC was supported by CAPES Foundation Grant No. 18668127; Instituto Gulbenkian de Ciência (IGC); and Indiana University Precision Health to Population Health (P2P) Study.
LMR was partially funded by the National Institutes of Health, National Library of Medicine Program, grant 01LM011945-01, by a Fulbright Commission fellowship, and by NSF-NRT grant 1735095 ``Interdisciplinary Training in Complex Networks and Systems.''
The funders had no role in study design, data collection and analysis, decision to publish, or preparation of the manuscript.

\section*{Acknowledgments}
We would like to thank Manuel Marques-Pita, Santosh Manicka, and Etienne Nzabarushimana for helpful conversations throughout the development of the CANA package.

\section*{Data Availability Statement}
The CANA python package and all datasets analyzed for this study can be found on Github at \url{github.com/rionbr/CANA}.
% Please see the availability of data guidelines for more information, at https://www.frontiersin.org/about/author-guidelines#AvailabilityofData

\bibliographystyle{frontiersinSCNS_ENG_HUMS} % for Science, Engineering and Humanities and Social Sciences articles, for Humanities and Social Sciences articles please include page numbers in the in-text citations
\bibliography{CANAref}

%%% Make sure to upload the bib file along with the tex file and PDF
%%% Please see the test.bib file for some examples of references

\section*{Figure captions}

%%% Please be aware that for original research articles we only permit a combined number of 15 figures and tables, one figure with multiple subfigures will count as only one figure.
%%% Use this if adding the figures directly in the mansucript, if so, please remember to also upload the files when submitting your article
%%% There is no need for adding the file termination, as long as you indicate where the file is saved. In the examples below the files (logo1.eps and logos.eps) are in the Frontiers LaTeX folder
%%% If using *.tif files convert them to .jpg or .png
%%%  NB logo1.eps is required in the path in order to correctly compile front page header %%%

%%% If you are submitting a figure with subfigures please combine these into one image file with part labels integrated.
%%% If you don't add the figures in the LaTeX files, please upload them when submitting the article.
%%% Frontiers will add the figures at the end of the provisional pdf automatically
%%% The use of LaTeX coding to draw Diagrams/Figures/Structures should be avoided. They should be external callouts including graphics.

\end{document}